\documentclass[sigconf]{acmart}

\renewcommand\footnotetextcopyrightpermission[1]{}
\usepackage{microtype}
\usepackage{graphicx}
\usepackage{subfigure}
\usepackage{booktabs} 
\usepackage{hyperref}
\usepackage{fontawesome}
\usepackage{wasysym}
\usepackage{booktabs}
\usepackage{multirow} 
\usepackage{multicol} 
\usepackage{pifont}  
\newcommand{\semicheckmark}{%
\textcolor{black}{\ding{51}}{\footnotesize\textcolor{black}{\kern-0.78em\raise0.18ex\hbox{\ding{55}}}}
}
\usepackage{verbatimbox}
\usepackage{xspace}
\usepackage{makecell}
\usepackage{stfloats}
\usepackage{subfigure}
\makeatletter
\DeclareRobustCommand\onedot{\futurelet\@let@token\@onedot}
\def\@onedot{\ifx\@let@token.\else.\null\fi\xspace}
\def\eg{\emph{e.g}\onedot} 
\def\ie{\emph{i.e}\onedot} 
 
\def\etc{\emph{etc}\onedot} 
 
\def\etal{\emph{et al}\onedot}
\makeatother

\AtBeginDocument{%
  }

\setcopyright{acmcopyright}
\copyrightyear{2018}
\acmYear{2018}
\acmDOI{XXXXXXX.XXXXXXX}

\begin{document}

\title{Who is Speaking Actually? Robust and Versatile Speaker Traceability for Voice Conversion}
\newcommand*{\affaddr}[1]{#1} 
\newcommand*{\affmark}[1][*]{\textsuperscript{#1}}
\newcommand*{\emailmark}[1]{\texttt{#1}}

\author{
Yanzhen Ren\affmark[1], Hongcheng Zhu\affmark[1], Liming Zhai\affmark[2], Zongkun Sun\affmark[1], Rubing Shen\affmark[1], Lina Wang\affmark[1]\\
\affaddr{\affmark[1]School of Cyber Science and Engineering, Wuhan University, China}\\
\affaddr{\affmark[2]Nanyang Technological University, Singapore}\\
}

\renewcommand{\shortauthors}{Ren, Zhu, et al.}

\begin{abstract}
  Voice conversion (VC), as a voice style transfer technology, is becoming increasingly prevalent while raising serious concerns about its illegal use. Proactively tracing the origins of VC-generated speeches, \ie, speaker traceability, can prevent the misuse of VC, but unfortunately has not been extensively studied. In this paper, we are the first to investigate the speaker traceability for VC and propose a traceable VC framework named VoxTracer. Our VoxTracer is similar to but beyond the paradigm of audio watermarking. We first use unique speaker embedding to represent speaker identity. Then we design a VAE-Glow structure, in which the hiding process imperceptibly integrates the source speaker identity into the VC, and the tracing process accurately recovers the source speaker identity and even the source speech in spite of severe speech quality degradation. To address the speech mismatch between the hiding and tracing processes affected by different distortions, we also adopt an asynchronous training strategy to optimize the VAE-Glow models. The VoxTracer is versatile enough to be applied to arbitrary VC methods and popular audio coding standards. Extensive experiments demonstrate that the VoxTracer achieves not only high imperceptibility in hiding, but also nearly 100\% tracing accuracy against various types of audio lossy compressions (AAC, MP3, Opus and SILK) with a broad range of bitrates (16 kbps - 128 kbps) even in a very short time duration (0.74s). Our speech demo is available at \url{https://anonymous.4open.science/w/DEMOofVoxTracer/}.
\end{abstract}



\maketitle

\section{Introduction}
\label{Introduction}
Voice conversion (VC) is a technology that converts the voice of a source speaker to sound like that of a target speaker while preserving its linguistic content. With the blossom of deep learning, the VC has gained more and more attention and opens up a wide variety of applications including speaker anonymization \cite{Qian2019speech, srivastava2020evaluating}, movie dubbing \cite{mukhneri2020voice}, voice customization \cite{kanagawa2013speaker, latorre2014voice}, singing conversion \cite{luo2020singing}, \etc. Recently, web platforms and mobile Apps (\eg, MetaVoice \cite{MetaVoice2022}, Respeecher \cite{Respeecher2022} and Resemble AI \cite{Resemble2022}) for VC are constantly emerging, and they deliver converted speeches as products or even online service, making manipulating voices as easy as editing text. As a result, various types of voice-changing audios or videos are widely circulated and shared in social media (\eg, TikTok, YouTube and Twitter) and are also applied to voice calls and online meetings (\eg, Voicemod \cite{Voicemod2022}).

\begin{figure}[t]
    \centering
    \includegraphics[width=\columnwidth]{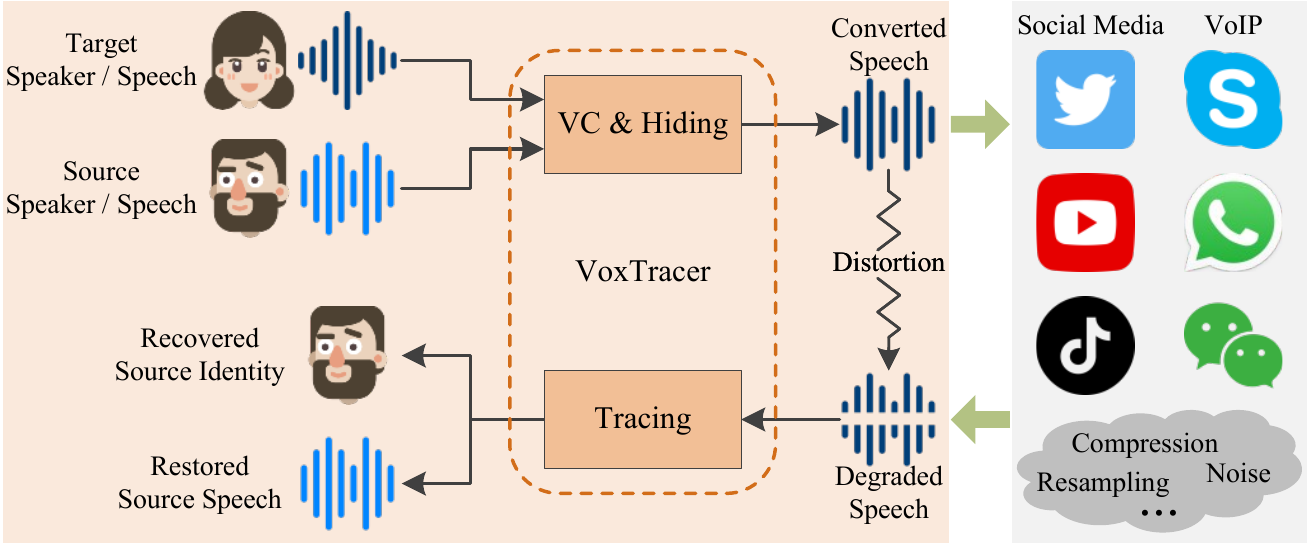}
    \caption{Speaker traceability for voice conversion (VC) in lossy environments. The waveform shape denotes the speech content while the waveform color denotes the speech timbre. The hiding process is parallel with VC, and the tracing process can accurately restore the source speaker identity and the source speech from the degraded speech.}
    \label{fig:VoxTracer}
\end{figure}

The VC is a kind of audio DeepFake technology \cite{muller2022human}, and it may also have a malicious, misleading and even destructive potential, causing serious individual or societal harm, including reputational damage, financial losses, the incitement of violence, news media manipulation and election rigging. In 2019, a British company claimed it was tricked by a phone call, which impersonated its CEO's voice, into wiring money to fraudsters \cite{fraud2019}.
In 2021, Zimbabwe vice president Mohad's voice was cloned by detractors angling to tarnish his reputation as a national political leader \cite{reputation2021}.

To counteract the risks and threats of DeepFakes, proactively tracing the source of DeepFakes (\ie, traceability) is more essential than passive detection \cite{yi2022add}. For this purpose, many governments around the world have issued laws and regulations to ensure the accountability and responsibility of DeepFake platforms. The US accountability act requires the DeepFakes to contain digital watermarks clearly identifying the altered audio or visual elements \cite{USact2022}. China recently proposed draft rules stipulating that DeepFake service providers shall ensure their synthesis content can identify itself and be traced \cite{Chinadraft2022}.

For the VC, the source speaker identity is a critical clue for tracing the origin of a converted speech (\ie, speaker traceability). However, all the major public VC platforms cannot yet fully meet the requirement of traceability due to the following two challenges. First, the converted speeches uploaded and disseminated to social media always undergo inevitable lossy compressions, which heavily affect the accuracy of VC traceability, especially for low-bitrate compressions. Second, different audio coding standards or encoders have different characteristics, and they are forced by social media or chosen by disseminators in various and even unpredictable ways, which further aggravates the difficulty of VC traceability. One may expect that traditional audio watermarking can be used for VC traceability. But unfortunately, existing audio watermarking has unsatisfactory robustness against lossy compressions, and is often limited to a few audio coding standards (see Table \ref{tab:task_cmp}). Therefore, it is of great importance and pressing to develop new methodologies for effective VC traceability. 

\begin{table}[t]
  \caption{Task comparison of voice conversion, audio watermarking and speaker traceability (our VoxTracer). Traceability is the ability to trace the ID of a source speaker after disseminating the converted speech. Robustness measures how the traceability tolerates the perturbations (\eg, lossy compression) during the speech dissemination. Versatility measures how the traceability adapts to different conditions (\eg, VC methods and audio coding standards).}
  \label{tab:task_cmp}
  \centering
  \resizebox{0.99\columnwidth}{!}
  {
  \begin{tabular}{cccc}
  \toprule
  & Traceability & Robustness & Versatility \\
  \midrule
  Voice Conversion & \ding{55} & N/A & N/A \\
  Audio Watermarking & \semicheckmark & \semicheckmark & \semicheckmark \\
  Our VoxTracer & \ding{51} & \ding{51} & \ding{51} \\
  \bottomrule
  \end{tabular}
  }
\end{table}

In this paper, we investigate the speaker traceability for VC in lossy environments, in which the source speaker identity and even the source speech can be accurately recovered for tracing when the converted speech suffers from quality degradation caused by various lossy compressions and audio processing, as shown in Figure \ref{fig:VoxTracer}.

Specifically, we propose a traceable VC framework named VoxTracer, which mainly involves two processes: the hiding process for integrating the speaker identity into VC in an imperceptible manner and the tracing process for recovering the speaker identity and even source speech from degraded speech. We first extract the unique speaker embedding to represent the speaker identity. We then design a VAE-Glow structure, for which the ``embedding-latent-speech'' data flow can be transformed bijectively for the hiding and tracing of speaker embedding. To solve the mismatch between the converted speech and the degraded speech caused by information loss, the hiding and tracing are trained asynchronously. Finally, the recovered speaker embedding is used for speaker verification and source speech restoration.

Our VoxTracer can be incorporated with arbitrary VC methods to empower them with traceability while not noticeably affecting the speech quality. We conduct a large-scale evaluation of VoxTracer, which clearly outperforms previous audio watermarking on speaker tracing accuracy. We also comprehensively test the robustness of VoxTracer on main-stream audio coding standards (AAC, MP3, Opus and SILK) with a broad range of bitrates (16 kbps - 128 kbps), and the results show that our VoxTracer can achieve nearly 100\% tracing accuracy at as low as 16 kbps by just using a 0.74s long speech. Moreover, the VoxTracer is independent of the internal parts of VC and thus is versatile enough to plug in any existing one-stage VC and two-stage VC. A comparison of our VoxTracer with existing works is shown in Table \ref{tab:task_cmp}.

The contributions of this paper are summarized as follows:
\begin{itemize}
  \item \textbf{New Problem.} We are the first to investigate and address the problem of speaker traceability for the VC task by verifying source speakers and restoring source speeches.
  \item \textbf{Novel methodology.} We propose a VAE-Glow structure trained in an asynchronous fashion, which effectively solves the mismatch between hiding and tracing.
  \item \textbf{Strong robustness.} Our VoxTracer can achieve superior tracing accuracy against audio lossy compressions even at a very low bitrate.
  \item \textbf{High versatility.} Our VoxTracer is highly flexible and can be used for arbitrary VC methods and popular audio coding standards.
\end{itemize}

\section{Related Work}
\subsection{Voice Conversion (VC)}
Deep learning has dominated the recent studies of VC, which combines the content representations disentangled from source speeches with target timbre information to generate converted results \cite{ContentVec}, such as AutoVC \cite{autovc}, SIG-VC \cite{zhang2022sig}, S3PRL-VC \cite{huang2022s3prl}, disentangled sequential variational autoencoder (DSVAE) \cite {lian2022robust}, YourTTS-VC \cite{casanova2022yourtts}, AVQVC \cite{tang2022avqvc}, \etc. Some latest methods \cite{SpeechSplit2, autopst, QianZCHC20} further refine and disentangle the speech to get more diverse speaker-dependent information, which improves the conversion quality and controllability.

The deep learning-based VC can be roughly divided into two categories. The \emph{first category} is the mainstream two-stage VC that first generates Mel-spectrograms, which are then converted into speech waveforms by a pre-trained vocoder \cite{autovc, SpeechSplit2, zhang2022sig, huang2022s3prl, lian2022robust, autopst, QianZCHC20}. The \emph{second category} is one-stage VC \cite{casanova2022yourtts,tang2022avqvc} that directly generates speech waveforms using an end-to-end learning scheme .

However, none of the existing VC methods can guarantee that the converted speeches will not be used for illegal purposes, and none of them have feasible solutions to address the untraceability problem of converted speeches. Our VoxTracer can solve this problem by empowering the VC with speaker traceability and can be applied to arbitrary VC methods (including two-stage VC and one-stage VC).

\subsection{Audio Watermarking}
Traditional audio watermarking embeds watermarks in time domain or transform domain. Typical time domain methods \cite{Unoki2011EmbeddingLW, YongXiang2012ADT} are of simple implementation but are beset by their limited robustness. The transform domain methods include spread spectrum (SS-SNR-HS) \cite{SS-SNR-HS}, singular value decomposition \cite{SVD}, patchwork techniques \cite{patch}, \etc, which have better robustness but at a high computational expense.

Recent studies start using deep learning for audio watermarking. Pavlović \etal \cite{pavlovic2022robust} proposed a deep neural network based robust speech watermarking method (DNN-RSW) by jointly optimizing an embedder and a detector. Liu \etal \cite{liu2022dear} introduced a distortion layer into the training of watermarking models for enhancing robustness against re-recording.

For the application to speaker traceability, our VoxTracer has the following advantages over audio watermarking: 
1) The audio watermarking hides error-prone watermarks into generated speeches after the VC is performed. In contrast, our VoxTracer directly integrates the speaker embedding which is less interfered with by recovery error into the VC process in a unified manner. 
2) Our VoxTracer can further restore the original source speech for a higher grade of traceability, which is not possessed by audio watermarking. 
3) The audio watermarking has limited robustness against audio compression, while our VoxTracer can resist various types of lossy compressions with even low bitrates.

\begin{figure*}[t]
  \centering
  \includegraphics[width=\textwidth]{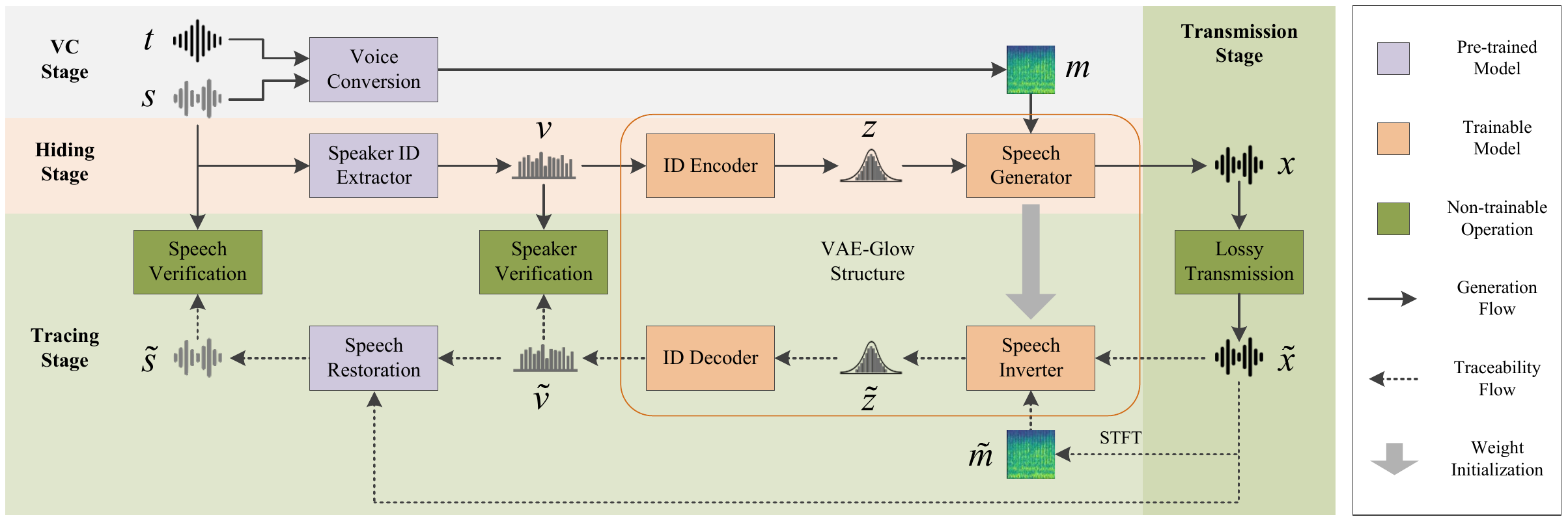}
  \vspace{-20pt}
  \caption{The framework of VoxTracer. The VC stage produces a converted Mel-spectrogram, with which the hiding stage integrates the source speaker identity into the generation of speech waveform. The transmission stage perturbs the converted speech. The tracing stage recovers the source speaker identity and restores the source speech.}
  \label{fig:framework}
\end{figure*}

\section{Methodology} \label{sec:method}
\subsection{Overview} \label{subsec:overview}
Let us suppose the life cycle of the traceable VC is as follows. The VC platform provides an online VC service, and the users have to first register their voiceprints that will be used for later speaker verification. When a user enjoys the VC service, the speaker identity will be parallelly hidden in the converted speech without affecting the user experience. The converted speech may be transmitted on the Internet, during which it may also be lossily compressed or processed. If the converted speech is used for illegal purposes, the VC platform can trace the source speaker by recovering the speaker identity and even restoring the source speech.

According to the above life cycle, our VoxTracer is built upon four stages: VC stage, hiding stage, transmission stage and tracing stage. The framework of the VoxTracer is shown in Figure \ref{fig:framework} and the related notations are listed in Table \ref{tab:notations}.

In the VC stage, a pre-trained VC model accepts as input a source speech $s$ and a target speech $t$, and outputs a Mel-spectrogram $m$, which will be used for further speech generation. See Section \ref{subsec:vc_stage} for the details of the VC stage.

The hiding stage aims to integrate the speaker identity into the speech generation. Specifically, a pre-trained speaker ID extractor takes the source speech $s$ as input and produces a speaker embedding $v$ that represents the source speaker identity. Then an ID encoder maps the speaker embedding $v$ to a latent code $z$ following a Gaussian distribution, which is appropriate for subsequent speech generation. Finally, the latent code $z$ is fed to a speech generator conditioned on the Mel-spectrogram $m$ to synthesize a converted speech $x$ that contains source speaker identity. The details of the hiding stage are in Section \ref{subsec:hiding_stage}.

In the transmission stage, the converted speech $x$ is transmitted on the Internet and suffers from various lossy compressions and audio processing operations, thus degrading the quality of converted speech $x$ and resulting in a degraded speech $\tilde x$. See Section \ref{subsec:trans_stage} for the details.

The tracing stage is the inversion of the hiding stage, attempting to recover the speaker identity hidden in the degraded speech. Specifically, a speech inverter retransforms the degraded speech $\tilde x$ to a recovered latent code $\tilde z$ conditioned on a recovered Mel-spectrogram $\tilde m$ obtained from $\tilde x$. Then an ID decoder remaps the $\tilde z$ to a recovered speaker embedding $\tilde v$. Finally, the $\tilde v$ is used for two types of verification: speaker verification and speech verification. For the speaker verification, the $\tilde v$ is compared with all speaker embeddings extracted from registered speeches to determine the source speaker embedding $v$. For the speech verification, the $\tilde v$ and the converted speech $\tilde x$ are fed to a speech restoration model to obtain a restored speech $\tilde s$, which sounds exactly the same as the source speech $s$. The tracing details are in Section \ref{subsec:tracing_stage}.

\begin{table}[t]
  \vspace{-6pt}
  \caption{Notations and their corresponding meanings.}
  \label{tab:notations}
  \centering
  \resizebox{\linewidth}{!}{
  \begin{tabular}{cl}
    \toprule
    Notation & Meaning  \\
    \midrule
    $t$ & Target speech \\
    $s$ / $\tilde s$ & Source speech / Restored speech \\
    $x$ / $\tilde x$ & Converted speech / Degraded speech \\
    $m$ / $\tilde m$ & Mel-spectrogram / Recovered Mel-spectrogram \\
    $v$ / $\tilde v$ & Speaker embedding / Recovered speaker embedding \\
    $z$ / $\tilde z$ & Latent code / Recovered latent code \\
    \bottomrule
    \end{tabular}
  }
\end{table}

\subsection{Voice Conversion Stage} \label{subsec:vc_stage}
Our VoxTracer aims to be applied to arbitrary VC methods and empower them with speaker traceability.

To ensure the versatility of VoxTracer, it is better to avoid interfering with the internal parts of VC and regard the VC as a black box. Considering the fact that the Mel-spectrogram is the intermediate output of most advanced VC methods (\ie, two-stage VC), we use the Mel-spectrogram combined with the speaker identity to generate the converted speech that is traceable, and this will make the hiding process more independent of the VC process.

The two-stage VC can be directly used in our VoxTracer, in which case the speech generator can act as a vocoder when hiding source speaker identity. For the one-stage VC, the converted speech waveform is transformed into Mel-spectrogram via a short-time Fourier transform (STFT), and then the speech generator is used to re-generate the speech waveform, as did for the two-stage VC. 
The STFT and speech re-generation are highly efficient and only slightly increase the VoxTracer's time cost which is almost negligible.

\vspace{-4pt}
\subsection{Hiding Stage} 
\label{subsec:hiding_stage}
\subsubsection{Speaker ID extractor}
To uniquely identify the source speaker, we adopt the classical automatic speaker verification (ASV) model \cite{ge2e} as a speaker ID extractor and use the extracted speaker embedding to represent the speaker identity.
The speaker ID extractor comprises a stack of three LSTM layers with 768 cells each and a final dense layer to output a 256-D vector (\ie, speaker embedding). In our implementation, the extractor is pre-trained with GE2E loss \cite{ge2e} on a voice search corpus containing 36M utterances from 18K U.S. English speakers \cite{jia2018transfer}.

There are two reasons for choosing speaker embedding to represent the source speaker identity. First, the speaker embedding is error-resilient. It does not require a complete and accurate recovery of speaker embedding in the tracing stage, and the variations of speaker embedding due to lossy compressions are also tolerable for speaker verification. This has been validated in ASV tasks \cite{bai2021speaker}, and is beneficial for improving the robustness of speaker traceability. Second, the speaker embedding is speaker-dependent. It contains rich source timbre information, which can be used to further restore the original source speech after the speaker embedding is revealed.

\vspace{-4pt}
\subsubsection{ID encoder and speech generator} 
Inspired by the recent image and audio steganography \cite{lu2021large, jing2021hinet, xu2022robust, DPAS}, we perform the hiding of speaker embedding by using a generative flow (Glow) model \cite{kingma2018glow}, which generates realistic objects from latent codes in an invertible way.
To apply the Glow to our hiding process, the latent code can be replaced by the mapped speaker embedding, the Mel-spectrogram can be used as a condition to guide the speech generation, and the invertibility of Glow can be used for further tracing.

The latent codes of Glow always follow a Gaussian prior distribution \cite{kingma2018glow}, while the distribution of speaker embeddings is usually uniform or not explicitly known \cite{wang2020understanding}, making it impossible to directly feed the speaker embeddings to the Glow model for speech generation. To solve this problem, we design an ID encoder that maps the speaker embeddings into Gaussian-distributed latent codes, which are then fed into a Glow based speech generator to synthesize speeches.

For the ID encoder, we implement it with the encoder part of the VAE \cite{VAE}. The reasons are as follows. First, the VAE imposes a Gaussian prior on the latent code to enable it to follow a Gaussian distribution, which meets the Gaussianity requirement of Glow. Second, the variational inference adopted in the VAE encoder makes the VAE decoder more robust to input changes \cite{VAE}, and it is beneficial to the speaker identity tracing in the information loss scenario where the recovered latent codes (input of VAE decoder) are different from the hidden latent codes (output of VAE encoder). The decoder part of the VAE is asynchronously used in the tracing stage, which will be presented later.

The ID encoder is trained using the variational objective of the standard VAE. Given a speaker embedding $v$, a standard Gaussian prior $p(z) \triangleq {\mathcal{N} } (z; 0,\mathbf{I})$ and a Gaussian posterior ${q}(z|v) \triangleq {\mathcal{N} }(z;{\mu _z},{\sigma _z^2}\mathbf{I})$ with learnable mean $\mu _z$ and variance $\sigma _z$. The loss function of the ID encoder is as follows:
\begin{equation}
  {{\mathcal{L} }_{KL}} = {{\mathcal{D} }_{KL}}({q}(z|v) \,\Vert\, {{p}(z)}) \text{,}
  \label{eq:kl_loss}
\end{equation}
where ${{\mathcal{D} }_{KL}}(q \,\Vert\, p)$ denotes the Kullback-Leibler (KL) divergence between distributions $q$ and $p$.

For the Glow based speech generator, it is taken as a vocoder in two-stage VC to synthesize speeches from latent codes with converted Mel-spectrograms as its conditions. The latent code used in existing VC is randomly sampled and thus has no meaningful speech information, so we can hide the speaker identity into latent code without affecting the semantics of converted speeches. Specifically, we replace the random latent code with the mapped speaker embedding obtained by the ID encoder, and use the WaveGlow \cite{WaveGlow} as our speech generator.
Accordingly, the loss function of the speech generator is given by
\begin{equation}
  {{\mathcal{L} }_{gen}} = \log p(z) + \log \left| {\det (\mathrm{d}z / \mathrm{d}x)} \right| \text{,}
  \label{eq:gen_loss}
\end{equation}
where $\det (\mathrm{d}z / \mathrm{d}x)$ is the determinant of the Jacobian matrix.

In the hiding stage, the ID encoder and the speech generator are trained separately, since they are mutually independent and separate training can make the optimization easier.

\subsection{Transmission Stage} \label{subsec:trans_stage}
To ensure the VoxTracer's robustness in lossy environments, in this stage, we intentionally perturb the converted speeches via lossy compressions and audio processing operations for simulating real-world environments. Specifically, for the speech waveforms produced by the speech generator, we first save them in a WAV format, which will cause the saving error in the float-to-integer data conversion. Then, we compress the WAV-format speeches with a certain audio coding standard at a certain bitrate. We consider four coding standards including AAC, MP3, Opus and SILK. We also adopt common audio processing, such as noise attack, re-sampling, re-quantization, amplitude modification and filtering, to further test the robustness of VoxTracer. The detailed settings of lossy compressions and audio processing are in Section \ref{subsec:robustness} and Supplementary, respectively.

\subsection{Tracing Stage} 
\label{subsec:tracing_stage}
\subsubsection{Speech inverter and ID decoder}
We recover the source speaker identity by using two successive inverse models corresponding to those in the hiding stage.

The first model, \ie, speech inverter, shares the same network structure with the Glow based speech generator and is initialized with the weights of the speech generator for further training. Instead of directly using the speech-to-latent process of speech generator as speech inverter, fine-tuning the speech inverter can enhance the reconstruction of recovered latent code.
The reconstruction loss is used for training and is given by:

\begin{equation}
  {\mathcal {L}}_{z\_rec} = \left\|{\tilde{z}} - {z} \right\|_1 \text{,}
  \label{eq:z_rec_loss}
\end{equation}
where ${\left\| \cdot \right\|_1}$ denotes the L1 norm function.

The second model is the ID decoder, which is implemented with the decoder part of the VAE. The normal VAE decoder takes as input the random latent codes to improve robustness, so it can fit our information loss scenario where the recovered latent codes have random perturbations compared with the original ones.
Similar to Eq. (\ref{eq:z_rec_loss}), the loss function of the ID decoder is calculated as
\begin{equation}
  {\mathcal {L}}_{v\_rec} = \left\|{\tilde{v}} - {v} \right\|_1 \text{.}
  \label{eq:v_rec_loss}
\end{equation}

Different from the separate training in the hiding stage, the speech inverter and the ID decoder are trained jointly using loss
\begin{equation}
  {\mathcal {L}}_{t} = {\lambda _{z}} {\mathcal {L}}_{z\_rec} + {\lambda _{v}} {\mathcal {L}}_{v\_rec} \,\text{,}
  \label{eq:tracing_loss}
\end{equation}
where ${\lambda _{z}}$ and ${\lambda _{v}}$ are weight factors for balancing different loss terms. In our additional experiments, we tried separate training in the tracing stage, but its tracing accuracy is slightly lower than that of joint training. We also tried L2 loss for training, but the results are worse than those of L1 loss. The comparisons of different loss functions are provided in Supplementary.

Compared with normal VAE and Glow models, our VAE-Glow structure is characterized by asynchronous training. In our VoxTracer, the encoder of VAE (\ie, ID encoder) and the forward model of WaveGlow (\ie, speech generator) are cascaded to synthesize speeches in the hiding stage, while the decoder of VAE (\ie, ID decoder) and the backward model of WaveGlow (\ie, speech inverter) are cascaded to recover speaker identity in the tracing stage. Unlike normal VAE and WaveGlow whose models are trained synchronously, the models in the hiding stage and the models in the tracing stages are trained separately (\ie, asynchronous training). This can focus the learning more specifically on the tracing, which helps address the mismatch between converted speech and degraded speech, thus enhancing the tracing robustness in lossy environments. We also tried synchronously training the hiding-stage models and tracing-stage models, but they cannot converge due to the aforementioned mismatch.

\subsubsection{Speaker verification} 
We adopt the cosine similarity and decision threshold which are commonly used in ASV tasks to decide whether the recovered speaker identity matches the speaker identity previously registered.

Specifically, in the training phase, we follow \cite{ai2022deepfake} to tune the threshold of cosine similarity and search for the optimal one to trade-off between the false rejection rate (FRR) and the false acceptance rate (FAR). In the testing phase, we calculate the cosine similarity between the recovered speaker embedding and the speaker embeddings from all registered speakers, and then choose the largest cosine similarity. If the chosen cosine similarity is larger than the threshold, the corresponding registered speaker is confirmed as the source speaker of the degraded speech.

\begin{figure*}[t]
  \centering
  \includegraphics[width=\textwidth]{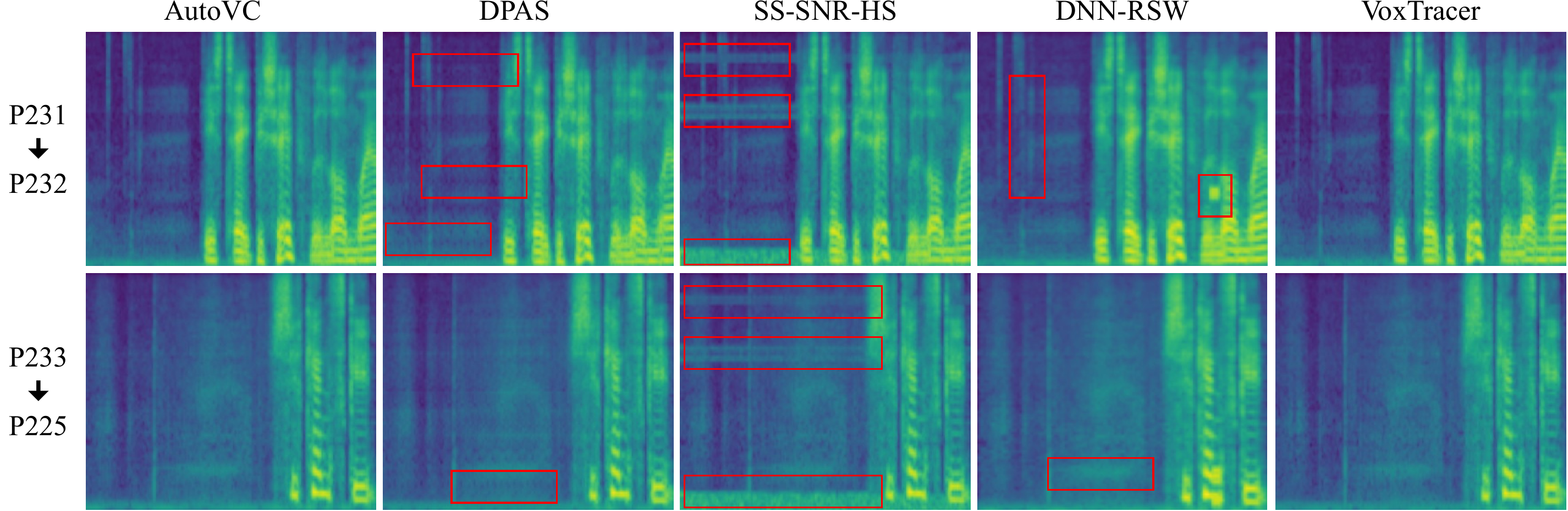}
  \vspace{-20pt}
  \caption{Visualization of Mel-spectrograms. The texts on the left are the conversion pairs. The hiding operations of DPAS, SS-SNR-HS, and DNN-RSW produce visible artifacts while VoxTracer does not.
  }
  \label{fig:mel_vis}
\end{figure*}

\subsubsection{Speech verification} 
To achieve full speaker traceability for judicial purposes, it is necessary to restore the source speech from the degraded speech rather than just verify the presence of the source speaker identity, and we refer to this process as speech verification. 

In the VC pipeline, the disentangled speaker embedding and content representation can be fused to synthesize a target speech. This motivates us to use the VC model to realize speech restoration. More concretely, we adopt the AutoVC \cite{autovc} as our speech restoration model. The degraded speech is fed to the content encoder of AutoVC to produce a content representation, which is combined with the recovered speaker embedding and fed to the decoder of AutoVC to produce the restored speech.

\section{Experiment}
\label{sec:experiment}

\subsection{Experimental Setup} 
\label{subsec:setup}
\subsubsection{Dataset} 
We use two datasets, VCTK Corpus \cite{vctk} and LibriSpeech Corpus \cite{librispeech}, for experiments. For the VCTK, we randomly select 100 speakers as source speakers and 10 speakers as target speakers. Each speaker has 10 sentences, resulting in 10,000 converted speeches, where 9,000 utterances are used for training and the rest for testing. For the LibriSpeech, we randomly select 2,000 source speakers and 4 target speakers with 5 sentences each to synthesize 40,000 utterances, and partition them into training and test sets by 4:1.

\subsubsection{Baseline} 
We adopt two types of baselines, one is audio steganography DPAS \cite{DPAS}, the other is audio watermarking including a traditional robust method SS-SNR-HS \cite{SS-SNR-HS} and a DNN-based robust speech watermarking method (DNN-RSW) \cite{pavlovic2022robust}. For a fair comparison, all the baselines hide binarized speaker embeddings into speeches and use the same speaker verification approach as our VoxTracer.

\subsubsection{Evaluation metrics.} We employ the mean opinion score (MOS) to evaluate the speech quality, for which the subjects are not informed of the details of test speeches and are asked to rate the intelligibility (MOS-I) and speaker similarity (MOS-S) by using a 5-point scale. More details are provided in Supplementary.

We use three types of metrics to evaluate speaker traceability. The first metric is tracing accuracy (TA), which refers to the proportion of correctly verified speaker embeddings to all embeddings, and similar metrics have been adopted in audio watermarking \cite{Hua_Goh_Thing_2015, Wang_Yuan_Unoki_2020}. To verify whether the tracing method can achieve a good balance between accuracy and precision, we further adopt the second metric, equal error rate (EER), which is defined as the point where FAR equals FRR on the receiver operating characteristic (ROC) curve (note that $\text{TA} = 1 - \text{FAR}$). The third metric is mean cosine similarity (MCS) which represents the similarity between the recovered speaker embeddings and the original ones.

Finally, we use the mean time cost (MTC) to measure the efficiency of hiding process. We do not calculate the MTC for tracing process, because the tracing process does not require low latency as the hiding process does. But actually, the time spent on the tracing process of our VoxTracer is almost negligible.

\subsubsection{Implementation details} 
In our experiments, we use an STFT with a window length of 1024 and a hop size of 256. All the trainable models of VoxTracer are optimized by using Adam optimizer with beta (0.9, 0.999) and a learning rate of $10^{-4}$. The batch sizes for the ID encoder and the speech generator are 8 and 16, while for the speech inverter and the ID decoder are both 20. ${\lambda _{z}} = 1$ and ${\lambda _{v}} = 1$. All models are implemented in PyTorch with one RTX 3090Ti GPU, and the average training time on VCTK Corpus and LibriSpeech Corpus is 16h and 3 days 15h. The detailed network architectures are provided in Supplementary.

\subsection{Speech Quality Evaluation} 
\label{subsec:converted_quality}

To measure the effect of speaker identity hiding on the perceptual quality of converted speeches (\ie, imperceptibility), we apply the baselines and our VoxTracer to three typical VC methods (AutoVC \cite{autovc}, SpeechSplit2.0 \cite{SpeechSplit2} and YourTTS-VC \cite{casanova2022yourtts}), for which the first two are two-stage VC and the last is one-stage VC. 

The speech quality scores of different tracing methods for VC are reported in Table \ref{tab:speech_quality}, where ``w/o hiding’’ means the plain VC outputs without any hiding operation involved. We observe that for each group of comparison, the plain VC outputs have the best speech quality, since their speech generation processes or outputs are not interfered with by speaker identity hiding. Our VoxTracer clearly outperforms all the baselines, and its speech quality scores are quite close to those of plain VC outputs. This demonstrates that our VoxTracer can be applied to arbitrary VC (two-stage VC and one-stage VC) without noticeably affecting the hearing experience.

\begin{table}[t]
  \vspace{-6pt}
  \caption{Speech quality scores of different tracing methods.}
  \label{tab:speech_quality}
  \centering
  \resizebox{0.9\linewidth}{!}{
  \begin{tabular}{cccc}
  \toprule
   VC & Method & MOS-I $\uparrow$ & MOS-S $\uparrow$ \\
  \midrule
  \multirow{5}{*}{AutoVC} & w/o hiding & 4.30 & 3.81 \\
   & DPAS & 4.00 & \textbf{3.80} \\
   & SS-SNR-HS & 3.54 & 3.69 \\
   & DNN-RSW & 3.17 & 3.46 \\
   & VoxTracer & \textbf{4.20} & \textbf{3.80} \\
  \midrule
  \multirow{5}{*}{SpeechSplit2.0} & w/o hiding & 4.35 & 3.89 \\
   & DPAS & 3.90 & 3.76 \\
   & SS-SNR-HS & 3.46 & 3.70 \\
   & DNN-RSW & 3.37 & 3.74 \\
   & VoxTracer & \textbf{4.31} & \textbf{3.89} \\
  \midrule
   \multirow{5}{*}{YourTTS-VC} & w/o hiding & 4.41 & 4.04 \\
   & DPAS & 4.00 & 3.87 \\
   & SS-SNR-HS & 3.63 & 3.76 \\
   & DNN-RSW & 3.39 & 3.72 \\
   & VoxTracer & \textbf{4.39} & \textbf{3.94}\\
  \bottomrule
  \end{tabular}
  }
\end{table}

To further compare the speech quality, we visualize the Mel-spectrograms of converted speeches from different tracing methods in Figure \ref{fig:mel_vis} (we only take the AutoVC as an example to save on space). We can see that the Mel-spectrograms of AutoVC and VoxTracer are almost indistinguishable, while the Mel-spectrograms of other tracing methods have visible artifacts compared with AutoVC. This again demonstrates the high speech quality of our VoxTracer.

\begin{table}[t]
  \caption{Performance comparison of different tracing methods on traceability accuracy (TA), equal error rate (EER), mean cosine similarity (MCS) and mean time cost (MTC).}
  \centering
  \label{tab:traceability}
  \resizebox{\linewidth}{!}{
  \begin{tabular}{cccccc}
  \toprule
  Compression & Methods & TA $\uparrow$ & EER $\downarrow$ & MCS $\uparrow$ & MTC $\downarrow$ \\
  \midrule
  \multirow{4}{*}{\makecell[c]{AAC\\96 kbps}} & DPAS & 1.39$\%$ & 48.90$\%$ & 0.4461 & \textbf{0.27s} \\
  & SS-SNR-HS & \textbf{100.00}$\%$ & 2.30$\%$ & 0.9310 & 27.12s \\
  & DNN-RSW & 96.40$\%$ & 3.10$\%$ & 0.9861 & 0.32s \\
  & VoxTracer & \textbf{100.00}$\%$ & \textbf{0.00}$\%$ & \textbf{0.9997} & \textbf{0.27s}\\
  \midrule
  \multirow{4}{*}{\makecell[c]{AAC\\32 kbps}} & DPAS & 1.25$\%$ & 48.40$\%$ & 0.4732 & \textbf{0.27s} \\
  & SS-SNR-HS & 69.60$\%$ & 35.10$\%$ & 0.6669 & 27.85s \\
  & DNN-RSW & 96.50$\%$ & 2.70$\%$ & 0.9862 & 0.32s \\
  & VoxTracer & \textbf{100.00}$\%$ & \textbf{0.00}$\%$ & \textbf{0.9974} & \textbf{0.27s}\\
  \midrule

  \multirow{4}{*}{\makecell[c]{MP3\\96 kbps}} & DPAS & 1.63$\%$ & 46.10$\%$ & 0.4398 & \textbf{0.27s} \\
  & SS-SNR-HS & \textbf{100.00}$\%$ & \textbf{0.00}$\%$ & \textbf{0.9999} & 27.96s \\
  & DNN-RSW & 96.30$\%$ & 3.30$\%$ & 0.9856 & 0.32s \\
  & VoxTracer & \textbf{100.00}$\%$ & \textbf{0.00}$\%$ & 0.9997 & \textbf{0.27s}\\
  \midrule
  \multirow{4}{*}{\makecell[c]{MP3\\32 kbps}} & DPAS & 1.62$\%$ & 46.60$\%$ & 0.4728 & \textbf{0.27s} \\
  & SS-SNR-HS & 65.40$\%$ & 28.90$\%$ & 0.5928 & 29.07s \\
  & DNN-RSW & 95.60$\%$ & 3.60$\%$ & 0.9847 & 0.32s \\
  & VoxTracer & \textbf{100.00}$\%$ & \textbf{0.00}$\%$ & \textbf{0.9990} & \textbf{0.27s}\\
  \midrule

  \multirow{4}{*}{\makecell[c]{Opus\\48 kbps}} & DPAS & 0.87$\%$ & 50.80$\%$ & 0.5378 & \textbf{0.27s} \\
  & SS-SNR-HS & 80.40$\%$ & 23.70$\%$ & 0.6928 & 28.08s \\
  & DNN-RSW & 3.90$\%$ & 41.60$\%$ & 0.4326 & 0.32s \\
  & VoxTracer & \textbf{100.00}$\%$ & \textbf{0.00}$\%$ & \textbf{0.9954} & \textbf{0.27s}\\
  \midrule
  \multirow{4}{*}{\makecell[c]{Opus\\24 kbps}} & DPAS & 1.19$\%$ & 49.60$\%$ & 0.5333 & \textbf{0.27s} \\
  & SS-SNR-HS & 58.80$\%$ & 25.60$\%$ & 0.5529 & 28.47s \\
  & DNN-RSW & 4.10$\%$ & 43.00$\%$ & 0.4327 & 0.32s \\
  & VoxTracer & \textbf{100.00}$\%$ & \textbf{0.00}$\%$ & \textbf{0.9938} & \textbf{0.27s}\\
  \midrule

  \multirow{4}{*}{\makecell[c]{SILK\\32 kbps}} & DPAS & 0.91$\%$ & 52.10$\%$ & 0.4968 & \textbf{0.27s} \\
  & SS-SNR-HS & 1.00$\%$ & 47.10$\%$ & 0.5081 & 28.40s \\
  & DNN-RSW & 2.60$\%$ & 44.30$\%$ & 0.4184 & 0.32s \\
  & VoxTracer & \textbf{100.00}$\%$ & \textbf{0.00}$\%$ & \textbf{0.9917} & \textbf{0.27s}\\
  \midrule
  \multirow{4}{*}{\makecell[c]{SILK\\16 kbps}} & DPAS & 0.84$\%$ & 47.30$\%$ & 0.5165 & \textbf{0.27s} \\
  & SS-SNR-HS & 1.30$\%$ & 47.80$\%$ & 0.5045 & 27.94s \\
  & DNN-RSW & 2.70$\%$ & 50.00$\%$ & 0.4171 & 0.32s \\
  & VoxTracer & \textbf{100.00}$\%$ & \textbf{0.00}$\%$ & \textbf{0.9944} & \textbf{0.27s}\\     
  
  \bottomrule
  \end{tabular}
  }
\end{table}

\subsection{Traceability Evaluation} 
\label{subsec:traceability}
To measure the speaker traceability of different tracing methods, we comprehensively compare the TA, EER, MCS and MTC of baselines and VoxTracer for AutoVC. The experiments are conducted on VCTK Corpus compressed with four audio coding standards, each of which uses a high bitrate and a low bitrate. The speech length for all tracing methods is 6.66s, because it is the minimum length for SS-SNR-HS to balance hiding capacity and speech quality. The results are summarized in Table \ref{tab:traceability}.

For the TA, the audio steganography DPAS cannot resist any lossy compression with any bitrate, since the hiding in audio steganography is always very fragile. The traditional audio watermarking SS-SNR-HS yields high TA for AAC and MP3 with high-bitrate compression, but has low TA for other conditions. The neural audio watermarking DNN-RSW performs well for AAC and MP3 but cannot work for Opus and SILK. This indicates that audio watermarking has a limited range of applications. As a striking contrast, our VoxTracer achieves 100\% TA in all situations, demonstrating its excellent speaker traceability as well as versatility. The EER results are completely consistent with TA. The MCS results of VoxTracer are all nearly 1 (perfect match of two speaker identities), confirming again the superiority of our VoxTracer in speaker traceability.
 
Real-time performance is another factor that affects user experience. For a 6.66s long speech, the average execution time of AutoVC is 0.27s. We record the total time cost of VC and speaker identity hiding for all methods in the last column of Table \ref{tab:traceability}. We observe that our VoxTracer and DPAS tie for first place in the time comparison, and they do not incur extra time consumption to VC, because their hiding process can be parallel with the two-stage VC process to reduce delay. For the two audio watermarking methods, the DNN-RSW takes a slightly longer time, but the SS-SNR-HS increases the time cost by hundreds of times, which is not suitable for real-time speaker identity hiding. For the one-stage VC, the average time spent by additional STFT and speech re-generation is 0.08s and the total MTC is 0.35s, meaning that our VoxTracer can also be very efficient for one-stage VC.

\subsection{Robustness Evaluation} 
\label{subsec:robustness}
In Section \ref{subsec:traceability}, we have verified the traceability and real-time performance of VoxTracer. In this subsection, we further verify the robustness of VoxTracer in a live-streaming VC scenario, where the converted speeches are generated chunk-by-chunk. For this scenario, the speaker identity can be repeatedly hidden in each chunk and traced by combining the recovered speaker identities of all chunks in a speech. More specifically, we assume there are $N$ ($N = 2k+1, k = 0, 1, 2, 3, \cdots$) chunks in a speech. Each chunk is hidden with a speaker embedding and the source speaker will be identified if no less than $\left\lfloor {{N \mathord{\left/{\vphantom {N 2}} \right. \kern-\nulldelimiterspace} 2}} \right\rfloor  + 1$ recovered speaker embeddings can be correctly verified. We perform this experiment on four audio coding standards with a broad range of bitrates: 
\begin{itemize}
  \item AAC: 128 kbps, 96 kbps, 64 kbps, 32 kbps and VBR.
  \item MP3: 128 kbps, 96 kbps, 64 kbps, 32 kbps and VBR.
  \item Opus: 64 kbps, 48 kbps, 32 kbps and 24 kbps.
  \item SILK: 40 kbps, 32 kbps, 25 kbps and 16 kbps.
\end{itemize}

\subsubsection{Robustness on VCTK corpus} 
We first run the experiment on VCTK corpus and show the results in the first row of Figure \ref{fig:robustness}. The tested speakers are seen in training, but the speeches for the same speaker in training and testing are non-overlapping. We observe that for all audio coding standards and all bitrates, the VoxTracer achieves tracing accuracy (TA) up to 99.87\% by just using one chunk (\ie, 0.74s for $N = 1$). With the increasing number of chunks, the tracing accuracy rapidly rises to 100\% in all cases. This indicates that our VoxTracer is sufficiently robust to various types of audio compression, and the repeated hiding of speaker identity is conducive to enhancing the robustness of speaker traceability.
 
\subsubsection{Robustness on LibriSpeech corpus} 
\begin{figure*}[t]
  \centering
  \includegraphics[width=\textwidth]{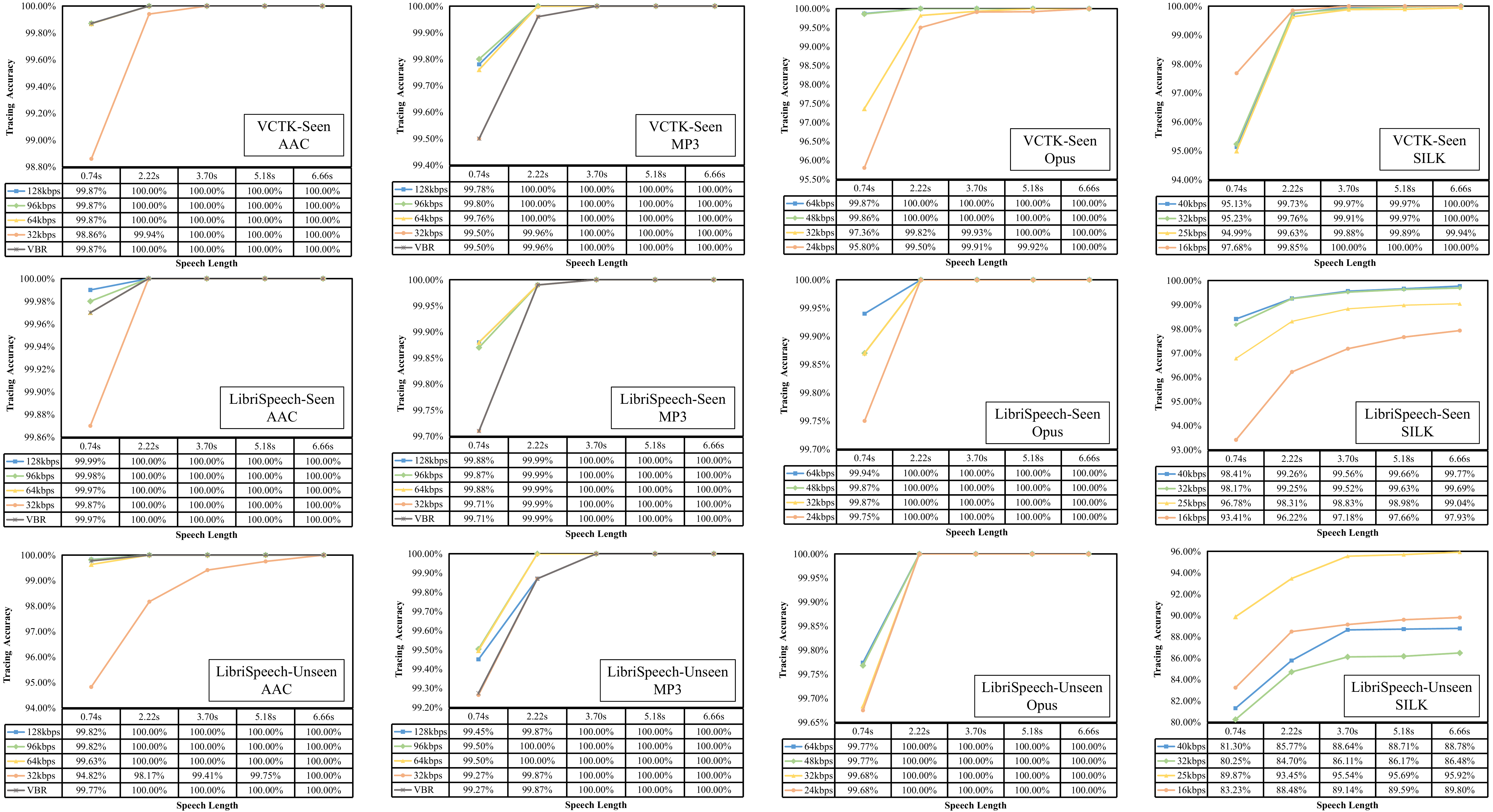}
  \caption{The tracing accuracy of VoxTracer as a function of speech length for different datasets, audio coding standards and bitrates. The experiments in the first and second rows are conducted in VCTK Corpus (100 seen speakers) and LibriSpeech Corpus (2000 seen speakers), respectively. The experiments in the third row are conducted in LibriSpeech Corpus (100 unseen speakers).}
  \vspace{-5pt}
  \label{fig:robustness}
\end{figure*}

We use another dataset LibriSpeech to evaluate the robustness in a large number of speakers. The configurations are the same as those on VCTK corpus, and the results are shown in the second row of Figure \ref{fig:robustness}. We can see that the results have the similar tendency with those on VCTK corpus, and the tracing accuracies are generally higher owing to training with more data, demonstrating that our VoxTracer can be applied to different datasets.

\subsubsection{Adaptability to unseen speakers} 
We evaluate the robustness of VoxTracer in a more challenging scenario where the tested speakers are unseen during training.
The experiment is performed by using 100 new speakers (each has three sentences) which are randomly selected from LibriSpeech corpus.
The results are shown in the third row of Figure \ref{fig:robustness}.
 
By comparing the last two rows in Figure \ref{fig:robustness}, we observed that the VoxTracer for unseen speakers can still maintain comparable performance for the AAC, MP3 and Opus, but its tracing accuracy drops for the SILK. This is because the SILK usually uses a low bitrate (maximum allowance up to 40 kbps), which easily causes severe speech quality degradation and thus affects the tracing robustness. But this problem can be solved by using a longer length of speech that contains more chunks to ensure the tracing accuracy.

\subsection{Analysis of Source Speaker Identity Leakage}
In the last two subsections, we have shown that our VoxTracer achieves excellent performance in speaker traceability even under severe distortion conditions. This raises a question: is the high tracing accuracy caused by source speaker identity leakage that the source speaker-related information is left in the converted speech?

To answer this question, we use the 10,000 converted speeches (generated by AutoVC \cite{autovc} ) of 100 source speakers from VCTK Corpus and adopt a pre-trained ASV model \cite{ge2e} to extract their speaker embeddings for speaker verification.
The results show that only 1.84\% of extracted embeddings are verified as their corresponding source speakers, which is nearly equal to random matching. This indicates that the VC does not suffer from source speaker identity leakage, and the high tracing accuracy only comes from our careful design, confirming the credibility of our tracing experiments.

\subsection{Quality Evaluation of Restored Speeches} 
\label{subsec:restored_quality}
As we have mentioned in Section \ref{subsec:tracing_stage}, we can even restore the source speeches for speech verification. To measure the similarity of the restored speeches and source speeches, we compare their speech quality scores with MOS-I and MOS-S and list the results in Table \ref{tab:restored_quality}.

For simplicity, we only use one type of VC method (AutoVC) to generate converted speeches and only use one bitrate (minimum bitrate) to compress the converted speeches. 

As seen from Table \ref{tab:restored_quality}, the restored speeches and the source speeches have close quality scores even at very low bitrates. Note that  the source speeches are not compressed, while the restored speeches are reconstructed from the degraded speeches that went through lowest-bitrate compressions. This indicates that the source speaker embeddings are recovered accurately so that they preserve the timbre information of source speakers. To further verify this conclusion, we also visualize the distributions of recovered speaker embeddings and source speaker embeddings in Supplementary to assess their similarity.

\begin{table}[h]
\caption{Quality comparisons of restored speeches and source speeches.}
\centering
\label{tab:restored_quality}
{
\vspace{-10pt}
\begin{tabular}{cccc}
  \toprule
  Speech Type & Compression & MOS-I $\uparrow$ & MOS-S $\uparrow$ \\
  \midrule
  Source Speeches & Uncompressed & 4.81 & 4.99 \\
  \midrule
  \multirow{4}{*}{Restored Speeches} 
  & AAC-32 kbps & 4.51 & 4.85 \\
  & MP3-32 kbps & 4.50 & 4.82 \\
  & Opus-24 kbps & 4.49  & 4.80 \\
  & SILK-16 kbps & 4.46  & 4.76 \\
  \bottomrule
\end{tabular}
}
\vspace{-10pt}
\end{table}

\section{Conclusion}
In this paper, we present VoxTracer, which is a traceable voice conversion (VC) framework aiming to empower the existing VC methods with speaker traceability. The VoxTracer builds upon an asynchronous-trained VAE-Glow structure that integrates the speaker identity into VC and recovers the speaker identity to verify the source speaker and restore the source speech. The VoxTracer is considerably robust against various audio processing operations and lossy compressions with low bitrates. Besides, the VoxTracer is highly versatile for arbitrary VC methods and mainstream audio coding standards. Last but not least, the VoxTracer is efficient and thus can be used for live-stream VC applications.

\bibliographystyle{ACM-Reference-Format}
\bibliography{voxtracer}

\appendix
\clearpage
\begin{figure*}[t]
  \centering
  \includegraphics[width=0.9\textwidth]{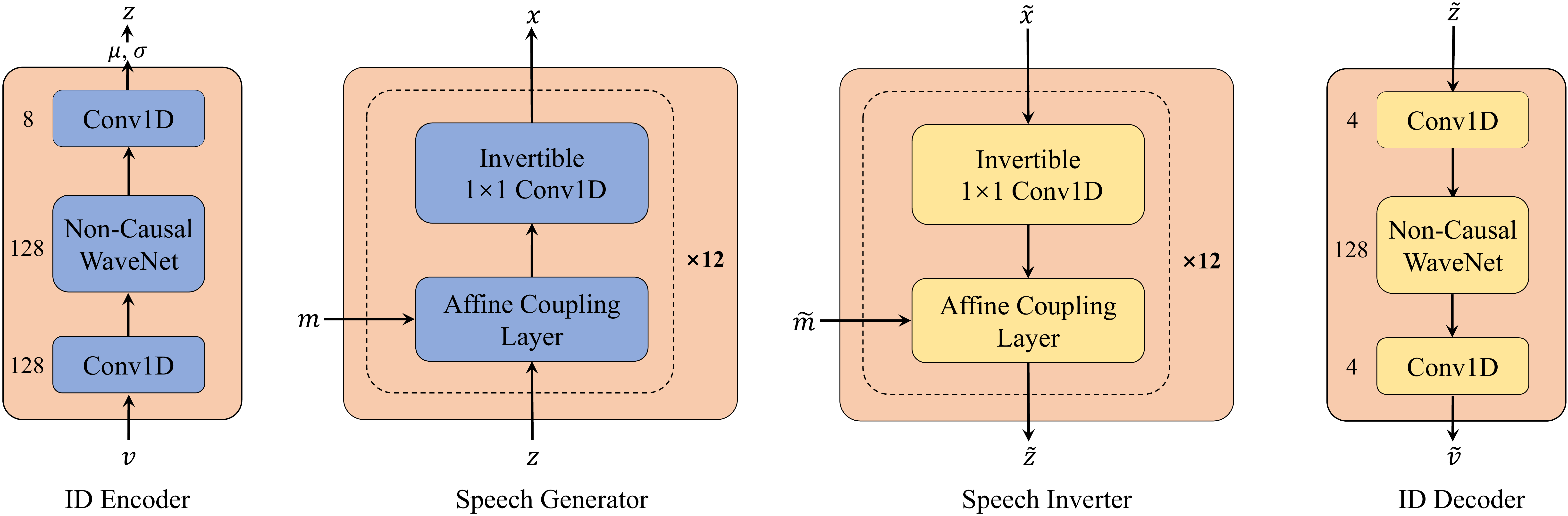}
  \caption{The network architecture of VoxTracer.}
  \label{fig:architecture}
\end{figure*}

\section{Network Architecture} 
\label{app:network}

We show the network architecture of our VoxTracer in Figure \ref{fig:architecture}. The ID encoder comprises a 1-D convolutional layer, a non-causal WaveNet and a 1-D convolutional layer. The first convolutional layer maps the input dimension from 4 up to 128, and the last layer maps the input dimension from 128 down to 8. The speech generator consists of 12 flow blocks, each of which has a $1 \times 1$ invertible convolutional layer and an affine coupling layer. The speech inverter shares the same architecture with the speech generator, and is initialized by the parameters of the speech generator. The ID decoder's architecture is similar to the latent encoder, except that the output dimension of the convolutional layer is 4 and the depth of non-causal WaveNet is halved.

\begin{table}[!htbp]
	\begin{center}
		\caption{The tracing accuracy of VoxTracer for different audio processing operations.}
		\label{tabaudio}
		\begin{tabular}{c | c | c}
			\toprule
			\multicolumn{2}{c|}{Processing Method} & Tracing Accuracy  \\
            \midrule
            \multirow{4}{*}{Guassian Noise} & 20dB & 99.79$\%$ \\
             & 30dB & 99.79$\%$ \\
             & 40dB & 99.82$\%$ \\
             & 50dB & 99.82$\%$ \\
             \midrule
             \multirow{2}{*}{Re-sampling} & 24000Hz & 99.79$\%$ \\
              & 16000Hz & 99.78$\%$ \\
             \midrule
             \multirow{2}{*}{Re-quantization} & 8bits & 99.70$\%$ \\
              & 32bits & 99.79$\%$ \\
             \midrule
             \multicolumn{2}{c|}{Amplitude Modification} & 99.78$\%$ \\
             \midrule
             \multicolumn{2}{c|}{Low-pass Filtering} & 99.50$\%$ \\
             \midrule
             \multicolumn{2}{c|}{Median Filtering} & 99.70$\%$ \\
			\bottomrule
		\end{tabular}
    \vspace{-5pt}
	\end{center}
\end{table}

\vspace{-5pt}
\section{Robustness against Audio Processing}  
\label{app:audio_processing}
To comprehensively evaluate the robustness of VoxTracer and keep consistent with the previous audio watermarking \cite{SS-SNR-HS, pavlovic2022robust, liu2022dear}, we  introduce distortions to converted speeches by using different audio processing operations as follows: 
\begin{itemize}
  \item Random Gaussian Noise. Add the random Gaussian noises with different signal-to-noise ratios (20 dB, 30 dB, 40 dB, 50 dB) to the converted speeches.
  \item Re-sampling. First sample the converted speeches from 22050 Hz down to 16000 Hz or up to 24000 Hz and then sample them back to 22050 Hz.
  \item Re-quantization. Re-quantize the converted speeches from 16 bits to 8 bits or 32 bits and then quantize them back to 16bits.
  \item Amplitude Modification. Decrease the amplitudes of converted speeches to 90$\%$ of their original values.
  \item Low-pass Filtering. Filter the converted speeches with the Butterworth filter at 1 kHz.
  \item Median Filtering. Filter the converted speeches with the median filter with a window size of 3.
\end{itemize}

These experiments are conducted on the test set (with a speech length of 0.74s) of VCTK, and the results are shown in Table \ref{tabaudio}. We can see that the VoxTracer achieves a tracing accuracy of no less than 99.50$\%$ in all situations, demonstrating that the VoxTracer is highly robust to various audio processing operations.

\begin{figure*}[t]
  \centering
  \includegraphics[width=2\columnwidth]{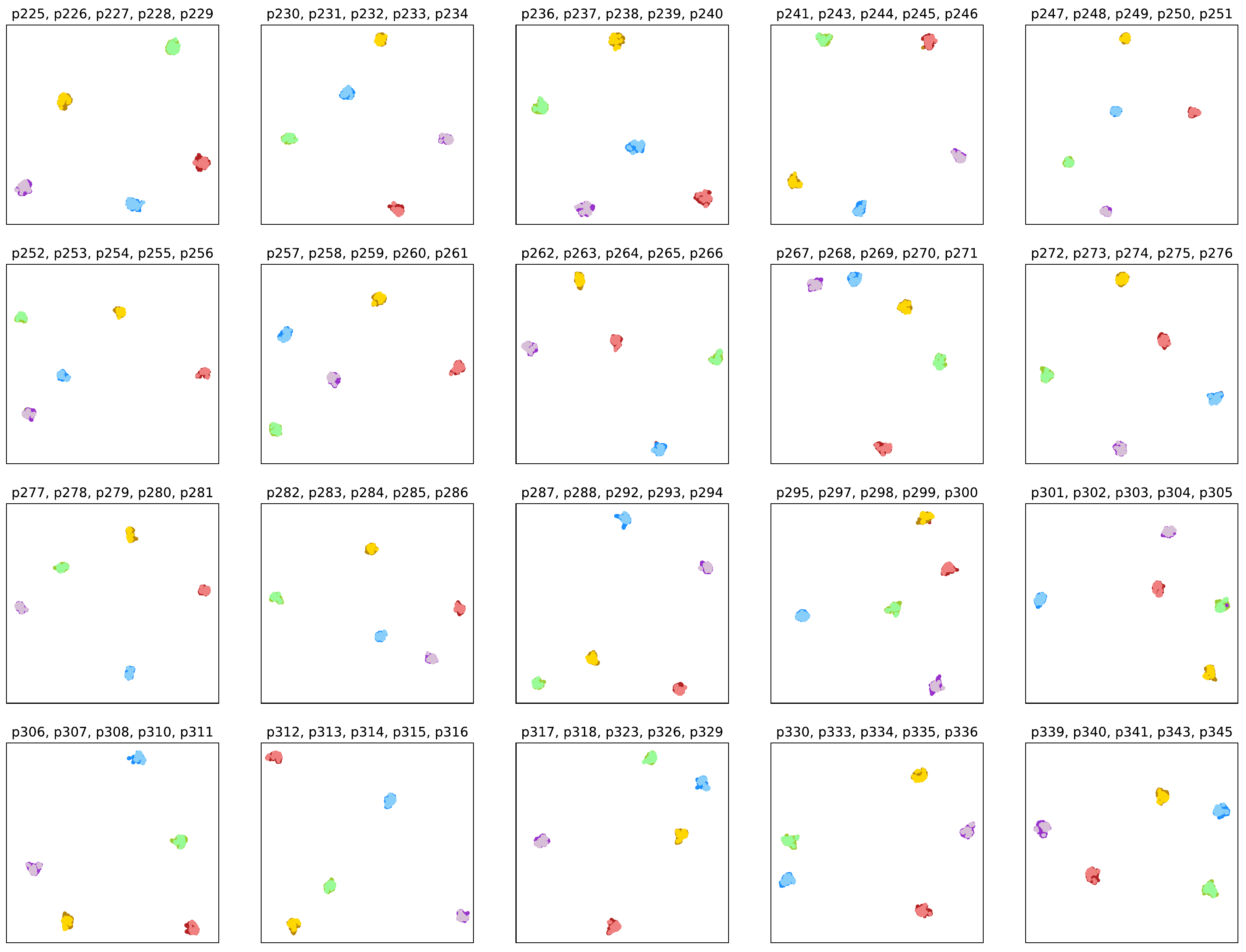}
  \caption{Visualization results of embedding distributions. Each subfigure contains 500 pair of embeddings for 5 speakers. Half of them are extracted from source speeches, and the other half are recovered by VoxTracer from degraded speeches. The title of each subfigure refers to the file ID of the 5 speakers in VCTK Corpus. In each subfigure, the overlapping color dots belong to the same speaker (the dark color denotes the source embeddings and the light color denotes the recovered embeddings).}
  \label{fig:vis_embedfing}
\end{figure*}

\section{Visualization of Recovered Speaker Embeddings}
\label{visionofdistribution}
To further measure the similarity of the restored speeches and source speeches, we compare the distributions of source embeddings and recovered embeddings. The former is extracted from the source speakers' real speeches and the latter is recovered by VoxTracer from degraded speeches compressed by AAC at 32 kbps.

We select 100 source speakers (each has 100 sentences)  from the VCTK Corpus, and divide them into 20 groups, each of which has 5 speakers. Then for each group, we obtain 500 pair of embeddings from source speeches and corresponding degraded speeches. Finally we visualize the 20 group of paired embeddings into 20 subfigures of Figure \ref{fig:vis_embedfing} by using UMAP \cite{umap}. 

As see from Figure \ref{fig:vis_embedfing}, the embedding distributions of the source speeches and degraded speeches for the same speaker overlap almost completely, while the embedding distributions for different speakers  exhibit clear distinctions. This explains why our VoxTracer can trace the source speakers accurately.

\begin{table}[!ht]
  \caption{Comparison of VoxTracer with different reconstruction loss, \ie, L1 norm (L1) and L2 norm (L2).}
  \centering
  \label{tab:ablation_loss}
  \begin{tabular}{ccc}
  \toprule
  Compression & Metric & Tracing Accuracy \\
  \midrule
  \multirow{2}{*}{\makecell[c]{AAC\\96 kbps}} & L1 & \textbf{99.87}$\%$ \\
   & L2 &  99.79$\%$ \\
  \midrule
  \multirow{2}{*}{\makecell[c]{AAC\\32 kbps}} & L1 & \textbf{98.86}$\%$ \\
   & L2 &  98.52$\%$ \\
   \midrule
  \multirow{2}{*}{\makecell[c]{Opus\\48 kbps}} & L1 & \textbf{99.86}$\%$ \\
   & L2 &  99.24$\%$ \\
  \midrule
  \multirow{2}{*}{\makecell[c]{Opus\\24 kbps}} & L1 & \textbf{95.80}$\%$ \\
   & L2 &  95.62$\%$ \\
  \bottomrule
  \end{tabular}
\end{table}

\begin{table}[t]
    \caption{Comparison of VoxTracer with different verification metrics, \ie, cosine similarity (cos) and L1 norm (L1).}
    \centering
    \label{tab:ablation_verif}
    \begin{tabular}{ccc}
    \toprule
    Compression & Metric & Tracing Accuracy \\
    \midrule
    \multirow{2}{*}{\makecell[c]{AAC\\96 kbps}} & cos & \textbf{99.87}$\%$ \\
     & L1 & 99.86$\%$ \\
    \midrule
    \multirow{2}{*}{\makecell[c]{AAC\\32 kbps}} & cos & \textbf{98.86}$\%$ \\
     & L1 & 98.30$\%$ \\
     \midrule
    \multirow{2}{*}{\makecell[c]{Opus\\48 kbps}} & cos & \textbf{99.86}$\%$ \\
     & L1 &  99.52$\%$ \\
    \midrule
    \multirow{2}{*}{\makecell[c]{Opus\\24 kbps}} & cos & \textbf{95.80}$\%$ \\
     & L1 &  95.46$\%$ \\
    \bottomrule
    \end{tabular}
\end{table}

\section{Selection of Different Loss Functions and Verification Metrics}
In Section 3.5 of our main paper, we use L1 norm to calculate the reconstruction loss in Eq. (3) and Eq. (4), but use cosine similarity for speaker verification. To measure the effect of different distance metrics on traceability, we build different variants of VoxTracer to compare their tracing accuracy. These experiments are conducted on the test set (with a speech length of 0.74s) of VCTK.

We first compare the reconstruction loss in Eq. (3) and Eq. (4) of our main paper with L1 norm and L2 norm and list the results in Table \ref{tab:ablation_loss}.  As can be seen, the L1 norm performs slightly better than the L2 norm, and that is why we select L1 norm to calculate the reconstruction loss.

We then report the tracing accuracy of VoxTracer using cosine similarity and L1 norm for speaker verification in Table \ref{tab:ablation_verif}. It is interestingly found that the cosine similarity based speaker verification achieves higher tracing accuracy than the L1 norm based version, even though the VoxTracer is optimized with L1 norm loss functions.

\begin{figure*}[!ht]
  \centering
  \subfigure{
      \begin{minipage}{7cm}
          \centering
          \includegraphics[width=\columnwidth]{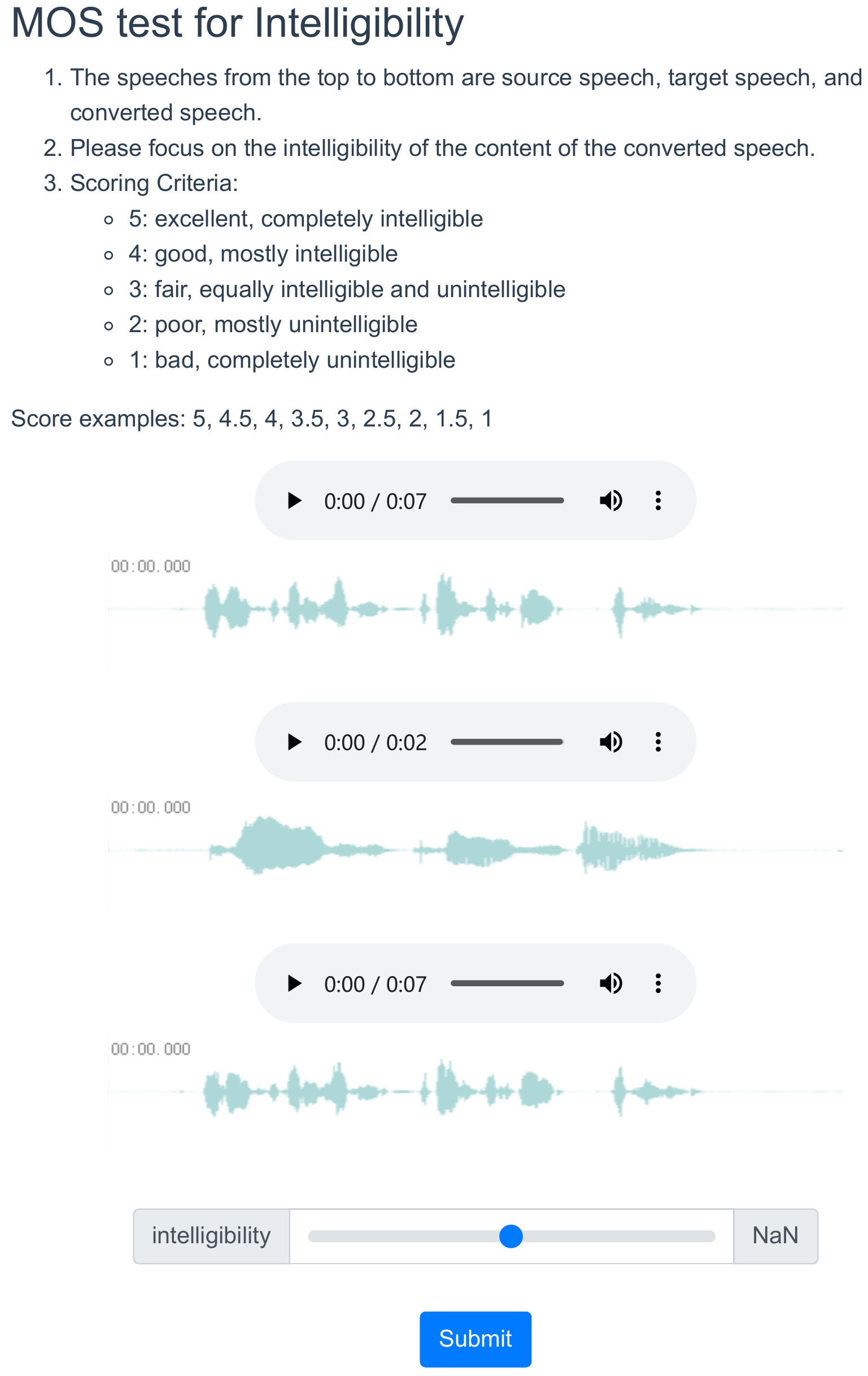}
      \end{minipage}    
  }
  \subfigure{
      \begin{minipage}{7cm}
          \centering
          \includegraphics[width=\columnwidth]{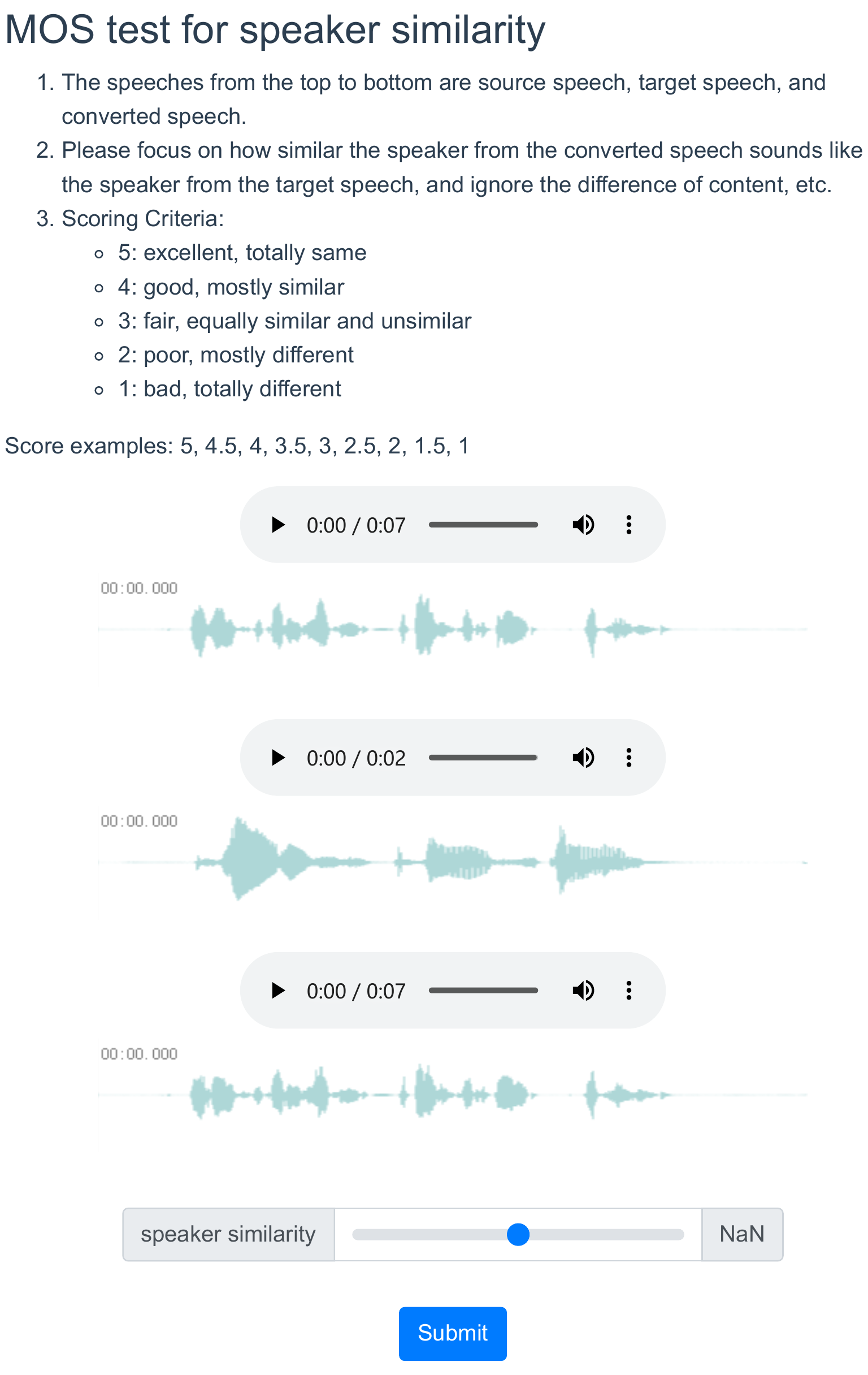}
      \end{minipage}    
  }
  \caption{The website screenshots of MOS test for intelligibility (left) and speaker similarity (right).}
  \label{fig:mos}
\end{figure*}

\section{MOS Test Details}
\label{MosDetails}
We built an online website for MOS test. The MOS test screenshots of intelligibility and speaker similarity are shown in Figure \ref{fig:mos}. We give the evaluation criteria on each website page, and the subjects are asked to follow these criteria strictly. Most importantly, all subjects are not informed of any information about the test speech in advance. In our implementation, for each MOS test, we select 10 subjects and present them with 50 test pairs randomly selected from the test set. We provide some examples in \url{https://anonymous.4open.science/w/DEMOofVoxTracer/}. We will open-source the code of VoxTracer upon acceptance of this manuscript.

\end{document}